\documentclass[prb,12pt]{revtex4}
\usepackage{amssymb,amsmath,natbib}
\usepackage{graphicx,subfigure,epsfig}
\usepackage{dcolumn}

\begin{document}

\title{\textbf{The Principle of Minimal Resistance in Non-Equilibrium Thermodynamics.}}

\author{\textbf{Roberto Mauri}}\email{r.mauri@ing.unipi.it}

\affiliation{Department of Civil and Industrial Engineering, Laboratory of Reactive Multiphase Flows, Universit\`a di Pisa, I-56126 Pisa, Italy; email: roberto.mauri@unipi.it\\}

\begin{abstract}

Analytical models describing the motion of colloidal particles in given velocity fields are presented. In addition to local approaches, leading to well known master equations such as the Langevin and the Fokker-Planck equations, a global description based on path integration is reviewed. This shows that under very broad conditions, during its evolution a dissipative system tends to minimize its energy dissipation in such a way to keep constant the Hamiltonian time rate, equal to the difference between the flux-based and the force-based Rayleigh dissipation functions. At steady state, the Hamiltonian time rate is maximized, leading to a minimum resistance principle. In the unsteady case, we consider the relaxation to equilibrium of harmonic oscillators and the motion of a Brownian particle in shear flow, obtaining results that coincide with the solution of the Fokker-Planck and the Langevin equations.

\end{abstract}

\maketitle

\section{Introduction} \label{sec I}

In classical thermodynamics, an isolated system tends to assume its stable equilibrium configuration which, according to the second law, is unique and corresponds to the state ${\mathbf x}$ in which the entropy functional $ S \left( {\bf x}\right) $ is maximized. As in classical thermodynamics any system has an infinite number of degrees of freedom, the equilibrium state is infinitely more probable than any other state. On the other hand, finite systems fluctuate, as all configurations have a finite probability to occur, equal to the Gibbs distribution, $C\exp \left[ S \left( {\mathbf x} \right) / k\right] $, where $k$ is Boltzmann's constant, and $C$ is a normalization factor.

Most of times, however, we deal with evolving dissipating systems, whose trajectories are described through appropriate master equations; in particular, we are interested either in systems that are maintained at steady state through imposed boundary conditions, or systems that are relaxing towards their state of stable equilibrium. Now, the evolution of conservative systems can be described either through a mechanistic approach, by integrating Newton's equations of motion, or through the variational principle of least action, leading to the Lagrangian and Hamiltonian formulations of classical mechanics. So, it is natural to look for a global description of irreversible processes as well, that is finding some universal function that could play the role of the action and the Lagrangian functionals in analytical mechanics. Naturally, just like the principle of least action and Lagrangian mechanics are equivalent to Newton's equation of motion, this novel variational approach should constitute a way to describe irreversible processes that is totally equivalent, although alternative, to solving the usual master equations.

The first example of a similar variational law is due to Kirchhoff \cite{Kirchhoff 1848}, who in the mid nineteenth century demonstrated that the steady state transport of electric charge obeys the principle of least dissipation of energy. This principle was generalized one hundred year later by Prigogine \cite{Prigogine 1961}, stating that under given boundary conditions, a system will tend to a state, in the eventual stationary process, that has a minimum of entropy production. De Groot and Mazur \cite{deGrootMazur 1962}  applied this principle to heat conduction, basically using the same approach as Kirchhoff' in electrostatics, stressing however that the principle of minimum entropy production is true only when the corresponding thermodynamic flux (i.e. the heat flux, $ {\mathbf J_Q} $ ) is proportional to its conjugated force (i.e. the gradient of the inverse temperature, $ \nabla T^{-1} $) through a phenomenological coefficient that is constant, i.e., independent of temperature \cite{note1}. A discussion about the history of this principle can be found in Jaynes \cite{Jaynes 1980} and in M\"{u}ller and Weiss \cite{MullerWeiss 2012}.

An apparently opposite principle was proposed by Ziegler \cite{Ziegler 1983}, stating that when the thermodynamic forces, $ \mathbf{F} $, are assigned, then the actual fluxes, $ \mathbf{J} $, maximize the entropy production rate, $ \dot{S} =  \mathbf{J} \cdot \mathbf{F} $. This principle generalizes the maximum (plastic) dissipation theorem of Mises, Taylor and Hill \cite{Hill 1950} to all nonequilibrium thermodynamics. In particular, when thermodynamic fluxes and forces are linearly related to one another, i.e. $ \mathbf{J} = \mathbf{L} \cdot \mathbf{F} $, with $ L_{ik} = L_{ki} $ denoting a constant generalized conductivity, then this theorem reduces to the well known principle, proposed by Onsager \cite{Onsager 1931b} in 1931, stating that when the thermodynamic forces, $ \mathbf{F} $, are assigned, the difference $ \dot{S} - \Psi_J $ between the entropy production rate and the flux-based Rayleigh's energy dissipation function, $ \Psi_J = \frac12 \left( \mathbf{J} \cdot \mathbf{L}^{-1} \cdot \mathbf{J} \right) $, is maximized. Vice versa, when thermodynamic fluxes are assigned, the difference $ \dot{S} - \Psi_F $ between the entropy production rate and the force-based Rayleigh's energy dissipation function, $ \Psi_F = \frac12 \left( \mathbf{F} \cdot \mathbf{L} \cdot \mathbf{F} \right) $, is maximized. Finally, for generic constraints, Onsager and Machlup \cite{OmL} and Gyarmati  \cite{Gyar70} found that  the difference $ \dot{S} - \Psi_J - \Psi_F $ between the entropy production rate and the sum of the two the energy dissipation functions, is maximized. This extremum property is generally referred to as the maximum entropy production (MEP) principle; its implications are reviewed in detail by Martyushev and Seleznev \cite{MartySel2006} \cite{Martyushev2013}  and by Verhas \cite{Verhas2014}, showing, in particular, how the parabolic equations appearing in heat and mass transport can be derived from it. Finally, Beretta \cite{Beretta2008}\cite{Beretta2014} extended the MET principle to far-from-equilibrium conditions, reformulating it as the steepest entropy ascent (SEA) model.

In this article, without trying to solve this fundamental problem in its entirety, we concentrate on a particular case, namely the time evolution of a fluctuating system, subjected to conservative forces. Apart from its simplicity, the great advantage of this problem is that for linear forces it reduces to the Ornstein-Uhlenbeck problem \cite{UO}, therefore providing a way to check the validity of our results. The novelty of the approach presented here is that we use the path integral formulation, thus providing an alternate point of view which might help to solve similar problems in the future.

\section{Description of the model} \label{sec II}

Consider a diffusing Brownian particle, subjected to a conservative force $ {\bf F} = - \nabla \phi_0 $, satisfying the Langevin equation,
\begin{equation}\label{1}
 \boldsymbol{\zeta} \cdot \left( \dot{\mathbf{x}} - \mathbf{V} \right)  = \mathbf{f},
\end{equation}
where the dot indicates time derivative, $ \boldsymbol{\zeta} $ is the resistance dyadic and we have defined the velocity $ \mathbf{V} = \mathbf{L} \cdot \mathbf{F} $, with $ \mathbf{L} = \boldsymbol{\zeta}^{-1} $ denoting a mobility tensor. Here, the Brownian force ${\bf f}$ results from the sum of a large number of collisions of the particle with the surrounding fluid, each occurring randomly and independently of the others, so that, according to the fluctuation-dissipation theorem, we obtain \cite{deGrootMazur 1962}:
\begin{equation} \label{2}
\left\langle \mathbf{f} \left( t \right) \right\rangle = \mathbf{0};
\hspace{2 em} \left\langle \mathbf{f} \left( t \right) \mathbf{f} \left( t
+ \tau \right) \right\rangle = 2 k T \, \boldsymbol{\zeta} \, \delta \left( \tau
\right).
\end{equation}

In general, using non-equilibrium thermodynamics notations, $ \dot{\mathbf{x}} $ can be identified with a generalized flux, $ \mathbf{J} $, while $  \boldsymbol{\zeta} \cdot \mathbf{V} $ is a generalized force, $ \mathbf{F} $; accordingly, $ \mathbf{L} = \boldsymbol{\zeta}^{-1} $ is the Onsager coefficient relating generalized fluxes and forces through the phenomenological linear relation, $ \mathbf{J} = \mathbf{L} \cdot \mathbf{F} $. Therefore, the generalized Langevin equation reads:
\begin{equation}\label{1A}
\mathbf{J} - \mathbf{L} \cdot \mathbf{F}  = \widetilde{\mathbf{J}},
\end{equation}
where $ \widetilde{\mathbf{J}} $ represents the fluctuating part of the flux, with,
\begin{equation} \label{2A}
\left\langle \widetilde{\mathbf{J}} \left( t \right) \right\rangle = \mathbf{0};
\hspace{2 em} \left\langle \widetilde{\mathbf{J}} \left( t \right) \widetilde{\mathbf{J}} \left( t
+ \tau \right) \right\rangle = 2 k T \mathbf{L} \delta \left( \tau \right).
\end{equation}

Note that, by applying the fluctuation-dissipation theorem, we have tacitly assumed that the generalized force is conservative, i.e. $ \mathbf{F} = - \nabla \phi_0 $ \cite{FDtheorem}.  For example, when we describe the motion of charged particles in electrostatics, $ \mathbf{J} $ is the electric current (equal to the mean particle velocity $ \dot{\mathbf{x}} $  per unit charge), while $ \mathbf{F} = - \nabla \phi_0 $ is the electric field induced by an electric potential, $ \phi_0 $; in the same way, $ \mathbf{J} $ can be a heat flux and $ \mathbf{F} = - \nabla T^{-1} $ the inverse temperature gradient.

Now, following the approach by Feynman and Hibbs \cite{FH} and applying the central limit theorem, we see that the probability of observing a certain Brownian force function ${\bf f} \left( t \right) $ consists of the following Gaussian distribution,
\begin{equation}\label{3}
\Pi \left[ {\bf f} \left( t \right) \right] \propto \exp \left[ -\frac
12 \iint \left[ \mathbf{f} \left( t \right) \cdot \mathbf{B} \left( \tau \right) \cdot \mathbf{f} \left( t + \tau \right) \right] dt d\tau \right],
\end{equation}
where $ {\bf B} $ is a sort of inverse of the variance of the process,
\begin{equation}\label{4}
\mathbf{B} \left( \tau \right) = \left\langle \mathbf{f} \left( t \right)
\mathbf{f} \left( t + \tau \right) \right\rangle^{-1} = \frac1 {2 k T} \boldsymbol{\zeta}^{-1} \delta \left( \tau \right).
\end{equation}

Equations (\ref{2}),  (\ref{3}) and (\ref{4}) reveals that, as the intensity of the fluctuating force is proportional to the local resistance, particles tend to diffuse from regions of high to regions of low fluctuations and, accordingly,  a system will tend to move along the paths of minimum resistance.

Now, since ${\bf f} \left( t \right) $ and ${\bf x}\left( t\right) $ are linearly related to one another through the Langevin equation (\ref{1}), the probability $ \Pi \left[ {\bf x} \left( t \right) \right] $ that the particle follows the path ${\bf x} \left( t \right) $ is proportional to $ \Pi \left[ {\bf f} \left( t \right) \right] $. Consequently, substituting Eqs. (\ref{1}) and (\ref{4}) into (\ref{3} ), we obtain:
\begin{equation}\label{5}
\Pi \left[ {\bf x} \left( t \right) \right] = G \left( {\bf x},t|{\bf x}_0\right)
\exp \left( - \frac 1{4 k T} \int\limits_{{\bf x} \left( t \right) }
\left( \dot{\mathbf{x}} - \mathbf{V} \right) \cdot \boldsymbol{\zeta} \cdot \left( \dot{\mathbf{x}} - \mathbf{V} \right)  dt \right),
\end{equation}
where the normalizing term $ G $ is the Jacobian \cite{G77}, depending only on the end points (see the end of Chapter \ref{sec III}).

Finally, the conditional probability $ \Pi \left( {\bf x},t|{\bf x}_0\right) $ that the particle moves from an initial state ${\bf x}_0$ at time $t=0$ to a final state ${\bf x}$ at time $t$ will be equal to the sum of the contributions (\ref{4}) of all paths connecting the two events, obtaining
\begin{equation} \label{7}
\Pi \left( {\bf x},t|{\bf x}_0 \right) = G \left( {\bf x},t|{\bf x}_0\right)  \int \exp \left[ -\frac{\mathcal{E}
\left[ {\bf x} \left( \tau \right) \right] }{2 k T} \right] {\cal D}
\left[ {\bf x} \left( \tau \right) \right],
\end{equation}
where the integral is taken over all paths (note that path integrals can be defined rigorously even when the paths $ \mathbf{x}(\tau) $ are not continuous functions \cite{GY60}) such that ${\bf x}\left(0 \right) = {\bf x}_0$ and ${\bf x}\left( t \right) ={\bf x}$, with,
\begin{equation}\label{8}
\mathcal{E} \left[ {\bf x} \left( t \right) \right] = \int\limits_0^t \mathcal{L}\left[
{\bf x} \left( \tau \right) ,\tau \right] d\tau ,
\end{equation}
and,
\begin{equation}\label{9}
\mathcal{L} \left[ {\bf x} \left( \tau \right) ,\tau \right] = \frac12 \left( \dot{\mathbf{x}} - \mathbf{V} \right) \cdot \boldsymbol{\zeta} \cdot \left( \dot{\mathbf{x}} - \mathbf{V} \right).
\end{equation}

In general, $\mathcal{L} \left[ {\bf x} \left( \tau \right),\tau \right] $ is (one half of) the rate of energy dissipation at time $\tau $ as the system follows the path $ \mathbf{x}\left( \tau \right) $, and $\mathcal{E} \left[ {\bf x}\left( t \right) \right] $ coincides with (one half of) the energy dissipated along that trajectory during the time interval $[ 0-t ]$. Due to the obvious analogy with analytical mechanics, we could also denote $\mathcal{L} $ by \emph{Lagrangian time rate} and $ \mathcal{E} $ by  \emph{action time rate}.

The exponential dependence of the path probability on the action time rate has been demonstrated directly by Wang and El Kaabouchiu \cite{Wang2014} through the direct simulation of the random motion of $ 10^9 $ Brownian particles in a linear force field.

In the particular case of linear velocity fields, the same result was obtained by Onsager and Machlup \cite{OmL}, starting from the Fokker-Planck equation and following Wiener's original derivation of path integration \cite{W}.   In fact, the rate of energy dissipation (\ref{9}) coincides, apart from the sign, with the Onsager-Machlup function,
\begin{equation}\label{10}
\mathcal{L} \left[ {\bf x} \left( \tau \right) ,\tau \right] = - \left( \dot{S} - \Psi_J - \Psi_F \right),
\end{equation}
where,
\begin{equation}\label{11}
\dot{S} =  \mathbf{J} \cdot \mathbf{F}; \hspace{1 em} \Psi_J = \tfrac12 \mathbf{J} \cdot \mathbf{L}^{-1} \cdot \mathbf{J}; \hspace{1 em} \Psi_F = \tfrac12 \mathbf{F} \cdot \mathbf{L} \cdot \mathbf{F},
\end{equation}
with $ \mathbf{J} = \dot{\mathbf{x}} $,  $ \, \mathbf{F} = \boldsymbol{\zeta} \cdot \mathbf{V} $ and $ \mathbf{L} = \boldsymbol{\zeta}^{-1} $ denoting the generalized flux, the generalized force and the Onsager coefficient, respectively.  Here, $ \dot{S} $ is the entropy production rate, while $ \Psi_J $ and $ \Psi_F $ are Rayleigh's flux-based and force-based dissipation potentials, depending respectively on the rate of change and on the state of the system.

Considering that 
\begin{equation}\label{add1}
 \int_0^t \dot{S} dt =  \int_0^t  \mathbf{F}  \cdot \dot{\mathbf{x}}   dt = - \int_0^t  \nabla \phi_0 \cdot d \mathbf{x}  = -  \Delta \phi = - \left( \phi(\mathbf{x}) - \phi(\mathbf{x}_0)  \right),
\end{equation}
we see that the action time rate (\ref{8}) can also be written in the following form,
\begin{equation}\label{add2}
\mathcal{E} \left[ {\bf x} \left( t \right) \right] = \Delta \left( \phi \right) + \int\limits_0^t \mathcal{L}' \left[
{\bf x} \left( \tau \right) ,\tau \right] d\tau ,
\end{equation}
where
\begin{equation}\label{add3}
\mathcal{L}' \left[ {\bf x} \left( \tau \right) ,\tau \right] = \frac12 \dot{\mathbf{x}} \cdot \boldsymbol{\zeta} \cdot \dot{\mathbf{x}} + \frac12 \mathbf{V} \cdot \boldsymbol{\zeta} \cdot  \mathbf{V}  =  \Psi_J + \Psi_F.
\end{equation}

It is worth observing that, for imaginary $ t = - i u $, with $u$ real, these equations just give Feynman's path integral representation of a probability amplitude \cite{FH}.

\section{Minimum path.}  \label{sec III}

Among all paths, let us denote by ${\bf y}\left( \tau \right) $ the one that minimizes $\mathcal{E}$. According to the Hamilton-Jacobi formalism
of classical mechanics, the \emph{momentum time rate} ${\bf p}$ along the minimum path can be defined as,
\begin{equation}\label{12}
\mathbf{p} = \left[ \frac{\partial \mathcal{L}}{\partial \dot{\mathbf{x}}} \right]_{{\bf x} = {\bf y}} = \boldsymbol{\zeta} \cdot \left[ \dot{\mathbf{y}} - \mathbf{V} \left( \mathbf{y} \right) \right].
\end{equation}
Clearly, $ \mathbf{p} $ has the units of a generalized force and, in fact, it coincides with the Brownian force $ \mathbf{f} $ along the minimum path, i.e., $ \mathbf{p} \equiv \mathbf{f}_{min} $, where the subscript "$min$" indicates that the quantity is defined along the minimum path.

Now, define the \emph{Hamiltonian time rate} $\mathcal{H}$ (in fact, $\mathcal{H}$ has the units of an energy per unit time) as $ \mathcal{H} = {\bf p} \cdot \dot{{\bf y}} - \mathcal{L}_{min} $, where $  \mathcal{L}_{min} $ is (one half of) the rate of energy dissipation along the minimum path, i.e.,
\begin{equation}\label{12A}
 \mathcal{L}_{min} = \mathcal{L} \left[ {\bf y} \left( \tau \right),\tau \right]  =  \frac12 \left( \dot{\mathbf{y}} - \mathbf{V} \right) \cdot \boldsymbol{\zeta} \cdot \left( \dot{\mathbf{y}} - \mathbf{V} \right).
\end{equation}
 Thus, we obtain:
\begin{equation} \label{13}
\mathcal{H} = \frac12 \left( \dot{\mathbf{y}} + \mathbf{V} \right) \cdot \boldsymbol{\zeta} \cdot \left( \dot{\mathbf{y}} - \mathbf{V} \right) = \frac12 \dot{\mathbf{y}} \cdot \boldsymbol{\zeta} \cdot \dot{\mathbf{y}} - \frac12 \mathbf{V} \cdot \boldsymbol{\zeta} \cdot \mathbf{V},
\end{equation}
where $ \mathbf{V} = \mathbf{V} \left( \mathbf{y} \right) $, that is, using Rayleigh's dissipation potentials defined in Eq. (\ref{11}),
\begin{equation} \label{13B}
\mathcal{H} = \left( \Psi_J - \Psi_F \right)_{min}.
\end{equation}

Note that, if we minimize the action time rate (\ref{add2}) by defining the momentum time rate $ \mathbf{p}' =  \boldsymbol{\zeta} \cdot \dot{\mathbf{y}} $,  at the end we would obtain again the Hamiltonian time rate (\ref{13B}).

Here we see that, when $t \rightarrow -t$, then $ \dot{\mathbf{y}} \rightarrow - \dot{\mathbf{y}} $, while $ \mathbf{V} (\mathbf{y}) \rightarrow \mathbf{V} (\mathbf{y}) $, showing that $ \mathcal{H} $ is time-invariant along the minimum path \cite{note2}.  Indeed, it is intriguing that the evolution of a dissipative system can be described in terms of the conservative motion of a pseudo-system, whose "energy" $ \mathcal{H} $ is constant.

Expressing $ \mathcal{H} $ in terms of the generalized momenta (\ref{12}), we have:
\begin{equation} \label{13A}
\mathcal{H} \left( \mathbf{y}, \mathbf{p} \right) = \mathbf{p} \cdot \left( \mathbf{V} + \frac12 \boldsymbol{\zeta}^{-1} \cdot \mathbf{p} \right).
\end{equation} 

Now, substituting the expression (\ref{13A}) into the the canonical equation, $ \dot{{\bf p}} = - \partial \mathcal{H}  / \partial {\bf y} $, the minimum path can be determined explicitly. In particular, in the linear case, when $ \boldsymbol{\zeta} $ is constant we obtain:
\begin{equation}\label{14}
\dot{{\bf p}} + \left( {\bf \nabla}{\bf V} \right) \cdot {\bf p} = 0.
\end{equation}
This is the Euler-Lagrange equation, obtained by minimizing the energy dissipated (\ref{8}). Note that, in the isotropic case, when $ \boldsymbol{\zeta} = \mathbf{I} $, Eq. (\ref{14}) can be rewritten in the following simple form,
\begin{equation}\label{15}
\ddot{{\bf y}} = {\bf \nabla} U + \dot{{\bf y}} \times {\bf B},
\end{equation}
where
\begin{equation}\label{16}
U = \frac 12 V^2; \hspace{10mm} {\bf B} = - {\bf \nabla} \times {\bf V}.
\end{equation}
Similar results were obtained by Wiegler,\cite{We} who studied the motion of Brownian particles in irrotational velocity fields, where $ \mathbf{B} = \mathbf{0} $. So, the minimum path describes the trajectory of a pseudo-particle of unit mass and unit electric charge immersed in an electric field $U$ and a magnetic field $ {\bf B}$.

Now, in general, the domain of integration of the path integral is composed of all paths whose distance from the minimum path is of order $\delta \sim kT / \tilde{\zeta} \tilde{V}$ or less, where $\tilde{\zeta}$ and $\tilde{V}$ are typical values of $ \boldsymbol{\zeta} $ and ${\bf V}$, respectively. Therefore, expressing any path ${\bf x}\left( \tau \right) $ as the "sum" of the minimum path ${\bf y}\left( \tau \right) $ and a "fluctuating" part $ \widetilde{{\bf x}} \left( \tau \right) $ \cite{S},
\begin{equation} \label{17}
{\bf x} \left( \tau \right) = {\bf y} \left( \tau \right) + \widetilde{{\bf x}} \left( \tau \right) ,
\end{equation}
where $ \widetilde{{\bf x}} \left( 0 \right) = \widetilde{{\bf x}} \left( t \right) = 0 $, then $\mathcal{E} \left[ {\bf x} \left( t \right) \right] $ can be expanded formally around ${\bf y} \left( \tau \right) $ as:
\begin{equation} \label{18}
\mathcal{E} \left[ {\bf x} \left( t \right) \right] = \mathcal{E}_{\min} + \frac 12 \widetilde{{\bf x}} \widetilde{{\bf x}} \mathbf{:} \left[ \frac{\partial^2 \mathcal{E}}{\partial \widetilde{{\bf x}} \, \partial \widetilde{{\bf x}} } \right]_{{\bf x} = {\bf y}}  + \ldots,
\end{equation}
with $ \mathcal{E}_{\min} = \mathcal{E}\left[ {\bf y} \left( t \right), t \right] $, where we have considered that the first order derivative is identically zero. Accordingly, we see that, if within distances of $ O \left( \delta \right) $ from the minimum path, ${\bf x}$ can be approximated as a linear function, then $\mathcal{E}$ is a quadratic functional, and therefore the expansion (\ref{18}) terminates after the second derivative, with the last term being a function of $ \widetilde{{\bf x}} $ only, and not of ${\bf y}$ \cite{S}.  Finally, substituting (\ref{18}) into (\ref{7})-(\ref{9}), with $ \mathbf{x}(\tau) = \mathbf{y}(\tau) $, we obtain:
\begin{equation}\label{19A}
\Pi \left( {\bf x},t|{\bf x}_0\right) = G \left( {\bf x},t|{\bf x}_0\right) \exp \left[ -\frac 1{2 k T}
\int \limits_0^t \mathcal{L}_{\min} \left( \tau \right) d\tau \right],
\end{equation}
where the normalization function $ G $ depends only on the end points.

Subjected to a posteriori verification, we have:
\begin{equation}\label{6}
G \left( {\bf x},t|{\bf x}_0\right) = \exp \left[ - \frac12 \int \limits_0^t
{\bf \nabla} \cdot \left( \dot{{\bf x}} + \mathbf{V} \right) dt \right].
\end{equation}

Therefore, substituting this expression into Eq. (\ref{19A}), we obtain:
\begin{equation}\label{19}
\Pi \left( {\bf x},t|{\bf x}_0\right) = \exp \left[ -\frac{\phi}{k T} \right];  \hspace{2 em}  \phi = \frac12 \mathcal{E}_{min}^{(t)} = \int \limits_0^t \mathcal{L}_{\min}^{(t)} \left( \tau \right) d\tau,
\end{equation}
with,
\begin{equation}\label{20}
\mathcal{L}_{min}^{(t)} = \frac12 \left( \dot{\mathbf{y}} - \mathbf{V} \right) \cdot \boldsymbol{\zeta} \cdot \left( \dot{\mathbf{y}} - \mathbf{V} \right)  + k T \nabla \cdot \left( \dot{{\bf y}} + \mathbf{V} \right),
\end{equation}
and $ \mathbf{y} (t) $ represents the trajectory (\ref{14}) along the minimum path.  Note that the last term in the RHS of Eq. (\ref{20}) does not enter the minimization process, as its contribution only depends on the end points.

This result shows that under very general conditions the path integral is determined exclusively by the minimum path (\ref{14}), determining the Boltzmann-like distribution (\ref{19}). When compatible with the end points, the minimum path is obviously $ \mathbf{\dot{y}} = \mathbf{V} $.

Note that here, since the force field is conservative, i.e., $ \mathbf{F} = \boldsymbol{\zeta} \cdot \mathbf{V} = - \nabla \phi_0 $, with $ \phi_0 = \phi_0 (\mathbf{x}) $ denoting a time-independent potential, we obtain: $ \dot{\mathbf{y}} \cdot \boldsymbol{\zeta} \cdot \mathbf{V} = - d\phi_0 / dt $. Then, the steady state, equilibrium probability distribution of unconstrained systems can be obtained considering the reverse minimum path, $ \mathbf{\dot{y}} = - \mathbf{V} $, assuming that initially the system is at equilibrium, i.e.,  $ \mathbf{x} \left( t = 0 \right) = \mathbf{x}_0 = \mathbf{0} $, with $ \phi_0 (\mathbf{0}) = 0 $, while $ \mathbf{x} \left( t \right) = \mathbf{x} $. At the end, since $ \mathcal{L}_{\min} = \mathbf{\dot{y}} \cdot \nabla \phi_0 = \dot{\phi_0} $, we obtain the usual Boltzmann distribution, $ \Pi ( \mathbf{x}) = W \exp{\left[ - \phi_0(\mathbf{x})/ k T \right] } $.

\section{The Fokker-Planck equation.}

Now, instead of keeping both the lower and the upper limit of (\ref{19}) fixed, let us compare the different values of $ \phi  $ corresponding to trajectories having the same initial configuration, but with variable final configurations, i.e. when $ \delta \mathbf{x} (0) = \mathbf{0} $ and $ \delta \mathbf{x} (t) = \delta \mathbf{x} $. In this case, instead of $ \delta \phi = 0 $, we obtain $ \delta \phi = \frac12 \mathbf{p} \cdot \delta \mathbf{x} $, and therefore,
\begin{equation}\label{22}
    \nabla \phi = \frac12 \mathbf{p}.
\end{equation}
In the same way, $ \phi $  can be seen as a function of time, i.e. we consider the trajectories starting from the same initial configuration and ending at the same final configuration, but at a different time. From Eqs. (\ref{8}) and (\ref{19}), we se that the total derivative of $ \phi $ is:
\begin{equation}\label{23}
    \frac{d \phi}{d t} =  \frac12 \mathcal{L}_{min}^{(t)}.
\end{equation}

Note that, as the upper limit of the trajectory is free, the full expression (\ref{20}) of the Lagrangian has to be considered.  
Therefore, expressing the action rate, $ \phi $, as a function of both spatial and temporal coordinates of the upper limit, i.e., $ \phi = \phi \left( \mathbf{x}, t \right) $, we find:
\begin{equation}\label{24}
    \frac{d \phi}{d t} =  \frac{\partial \phi}{\partial t} + \frac{\partial \phi} {\partial \mathbf{x}} \cdot \dot{\mathbf{x}} =  \frac{\partial \phi}{\partial t} + \frac12 \mathbf{p} \cdot \dot{\mathbf{x}},
\end{equation}
thus obtaining the following Hamilton-Jacobi equation,
\begin{equation}\label{25}
     \frac{\partial \phi}{\partial t} + \frac12 \mathcal{H}^{(t)} \left( \mathbf{x}, \nabla \phi, t \right) = 0.
\end{equation}
Here, $ \mathcal{H}^{(t)} $ is the total Hamiltonian time rate,
\begin{equation}\label{26}
    \mathcal{H}^{(t)} \left( \mathbf{x}, \mathbf{p}, t \right) = \mathbf{p} \cdot \dot{\mathbf{x}} - \mathcal{L}_{min}^{(t)} = \mathbf{p} \cdot \left( \frac12 \boldsymbol{\zeta}^{-1} \cdot \mathbf{p} + \mathbf{V} \right) - 2 kT \, \nabla \cdot \left( \frac12 \boldsymbol{\zeta}^{-1} \cdot \mathbf{p} + \mathbf{V}  \right),
\end{equation}
with $ \mathbf{p} = 2 \nabla \phi $.   Therefore, Eq. (\ref{25}) becomes,
\begin{equation}\label{27}
     \frac{\partial \phi}{\partial t} + \nabla \phi \cdot \left( \boldsymbol{\zeta}^{-1} \cdot \nabla \phi + \mathbf{V} \right) = kT \nabla \cdot \left( \boldsymbol{\zeta}^{-1} \cdot \nabla \phi + \mathbf{V}  \right).
\end{equation}

Now, considering that the conditional probability is related to the generalized potential, $ \phi $, through $ \Pi \propto \exp{\left( - \phi / kT \right)} $, it is easy to see that the Hamilton-Jacobi equation (\ref{27}) is identical to the Fokker-Planck equation,
\begin{equation} \label{28}
\frac{\partial \Pi}{\partial t} + \nabla \cdot \mathbf{J} = 0; \hspace{2 em} \mathbf{J} = \mathbf{V} \Pi - kT \boldsymbol{\zeta}^{-1} \cdot \nabla \Pi,
\end{equation}
where  $ \mathbf{J} $ is the probability flux.  Therefore, we see that the quadratic approximation (\ref{19}), i.e. assuming that the minimum path is the only trajectory that contributes the path integral, is equivalent to assuming that the Fokker-Planck equation is a valid approximation to describe the random process. This result justifies assuming the expression (\ref{6}) for the normalization term.

\section{Steady state.}  \label{sec IV}

In this section we consider a system that is kept in a state of non-equilibrium through a set of constraints about its generalized forces or fluxes. In this case, considering that the time integral in Eq. (\ref{8}) reduces to a simple product, i.e., $ \mathcal{E} = \mathcal{L} t $, the rule of least energy dissipation becomes:
\begin{equation}\label{30}
    \mathcal{L} \left( {\bf x}, \dot{\mathbf{x}} \right) = \frac12 \left[ \dot{\mathbf{x}} - \mathbf{V} \left( {\bf x} \right)  \right] \cdot \boldsymbol{\zeta} \cdot \left[ \dot{\mathbf{x}} - \mathbf{V} \left( {\bf x} \right) \right] = \min
\end{equation}
This leads to Onsager's principle,
\begin{equation}\label{31}
    - \mathcal{L} = \left( \dot{S} - \Psi_J - \Psi_F \right) = \max,
\end{equation}
where $ \dot{S} $, $ \Psi_J $ and $ \Psi_F $ are defined in (\ref{11}). In addition, considering that at steady state $ d \phi / dt = \partial \phi / \partial t = \frac12 \mathcal{L}_{min}^{(t)} $, Eq. (\ref{25}) becomes:
\begin{equation}\label{31A}
   \mathcal{L}_{min}^{(t)} + \mathcal{H}^{(t)} = 0,
\end{equation}
so that we obtain:
\begin{equation}\label{32}
    \mathbf{p} \cdot \dot{\mathbf{y}} = 0.
\end{equation}
Consequently, considering that [cfr. Eq. (\ref{12})] $ \mathbf{p} = \boldsymbol{\zeta} \cdot \left( \dot{\mathbf{y}} - \mathbf{V} \right) $, we see that $ \dot{S} = 2 \Psi_J $, so that  Eq. (\ref{31}) can be rewritten as:
\begin{equation}\label{33}
    \mathcal{H} = \left( \Psi_J - \Psi_F \right) = \max,
\end{equation}
where Eq. (\ref{13B}) has been considered.  This shows that at steady state a system tends to maximize its Hamiltonian time rate. The same result could be obtained from Eq. (\ref{13A}), (\ref{31}) and (\ref{31A}).

Equation (\ref{33}) expresses a principle of minimum resistance, revealing that if in a process the generalized forces, $ \mathbf{F} $, are fixed, then the fluxes, $ \mathbf{J} $, are maximized, while when the generalized fluxes  are fixed, then the forces are minimized. This confirms that, since from Eqs. (\ref{2}),  (\ref{3}) and (\ref{4}) the intensity of the fluctuating force is proportional to the local resistance, a system will tend to move along the paths of minimum resistance. So, for example, if the temperature (or concentration) difference between two regions is fixed, heat (or chemical species) will tend to flow across regions of low resistance (or large conductivity), thus maximizing the flux. In that case, therefore, the entropy production rate will be maximized and, in fact, this behavior satisfies the maximum flux principle. Alternatively, when the heat flux is fixed, the rule of minimum resistance leads to minimizing the temperature (or concentration) difference, so that entropy production rate will be minimized.

Studying a similar problem, Martiouchev and Seleznev \cite{MS00} showed that the morphologies selected during crystallization tend to maximize the entropy production. Similar results were obtained by Molin and Mauri \cite{MM08}, who simulated the spinodal decomposition of a binary mixture confined between two walls that are quenched below the critical temperature. This rule was  demonstrated in great generality by Dewar \cite{D05} using nonequilibrium statistical mechanics, showing that it applies to all cases where macroscopic fluxes are free to vary under imposed constraints. Similar results were obtained by Favretti \cite{Favretti2009}. Accordingly, despite its name, this law generalizes Onsager's principle of least energy dissipation \cite{Onsager 1931b} to cases where systems are far from equilibrium and, in fact, it agrees with the steepest entropy ascent (SEA) model developed by Beretta \cite{Beretta2014}.

\section{Unsteady cases.} \label{sec V}

\subsection{Relaxation to equilibrium.}

In this section we consider unconstrained systems that relax towards equilibrium. In this case, since at equilibrium $ \mathbf{p} = \mathbf{V} = \mathbf{0} $, then $ \mathcal{H} = 0 $. Therefore, considering that $ \mathcal{H} $ is constant along the minimum path, we conclude that during its relaxation towards equilibrium a system satisfies the following relation:
\begin{equation}\label{34}
    \mathcal{H} = 0 \hspace{2 em} \rightarrow \hspace{2 em} \Psi_J = \Psi_F,
\end{equation}
where Eq. (\ref{13B}) has been considered.

As a simple example of application, consider a typical Ornstein-Uhlenbeck process, where a Brownian particle, with drag coefficient $ \zeta_{ij} = \zeta \delta_{ij} $ and diffusivity $ D = k T / \zeta $, is immersed in a quiescent fluid and is subjected to a linear potential force, (i.e., it is a harmonic oscillator), attracting the particle towards the origin, with $ \mathbf{F} = \zeta \mathbf{V} = - \zeta \mathbf{M} \cdot \mathbf{x} $, where $ M_{ij} = M_{ji} $. Accordingly, we can choose a reference frame where the axes coincide with the principal directions of the $ \mathbf{M} $ matrix, so that $ M_{ij} = M_i \delta_{ij} $, with $ M_i > 0 $.

Now, the conditional probability function $ \Pi \left( {\bf x},t| \mathbf{0} \right) $  that describes the motion of this Brownian particle is given by Eqs. (\ref{19})-(\ref{20}). Here, the minimum path equation (\ref{15}) reduces to:
\begin{equation} \label{41}
\ddot{y}_i = M_i^2 y_i,
\end{equation}
which, coupled to the conditions $ y_i(0) = 0 $ and $ y_i (t) = x_i $, yields:
\begin{equation} \label{42}
y_i (\tau) = x_i \frac{\sinh{ (M_i \tau)}}{\sinh{(M_i t)}}.
\end{equation}
Substituting this result into Eq. (\ref{20}), i.e. $ \mathcal{L}_{min} = \zeta \left( \dot{\mathbf{y}} - \mathbf{V} \right)^2  $, we obtain:
\begin{equation} \label{43}
\mathcal{L}_{min} = \zeta \sum_i \frac{M_i^2 x_i^2}{\sinh^2{(M_i t)}} \left( \cosh{(M_i \tau)} + \sinh{(M_i \tau)} \right)^2 + C (\tau),
\end{equation}
where $ C (\tau) $  is an irrelevant time-dependent function, independent of the endpoints. Finally, from Eq. (\ref{19}) we find the following Gaussian distribution \cite{Mauri2013}:
\begin{equation} \label{44}
\Pi \left( {\bf x},t| \mathbf{0} \right) = W(t) \exp \left[ - \frac{1 }{4 D} \sum_i { M_i x_i^2 \left[ 1 + \coth(M_it)\right] }\right],
\end{equation}
where $ W(t) $ is a normalization factor, which is independent of the endpoints. In particular, for long times, $ t \gg M^{-1} $, this solution tends to the equilibrium distribution,
\begin{equation} \label{45}
\Pi \left( {\bf x} \right) = W \exp \left[ - \frac{ 1 }{2 kT} \sum_i { g_i x_i^2 } \right]
\end{equation}
where $ g_i = \zeta M_i  $ represents the spring constant along the $i$-th direction. This is the well-known Ornstein-Uhlenbeck solution \cite{UO}, obtained by solving the Fokker-Planck equation; clearly, this result is not surprising, since in Eqs. (\ref{27}) and (\ref{28}) we saw that the Fokker-Planck equation is equivalent to the Hamilton-Jacobi equation for the mininum path.

\section{Uniform velocity field and simple shear flow.} \label{sec VI}

In this section we want to show that the path integral approach can also be applied to cases where there is no steady state. The simplest case is the diffusion of a Brownian particle in a uniform velocity field, i.e., when $ \mathbf{V} $ is a constant vector.  In this case, the minimum path $ \mathbf{y}(\tau) $ satisfies the Euler-Lagrange equation (\ref{15}), $ \ddot{\mathbf{y}} = 0 $, with $\mathbf{y}(0) = \mathbf{0}$ and $ \mathbf{y}(t) = \mathbf{x} $, obtaining: $ \mathbf{y}(\tau) = \mathbf{x} \tau / t $. Thus, the minimum path corresponds to a trajectory with uniform velocity, which of course has nothing to do with the "real" velocity of a diffusing Brownian particle. Now, substituting this result into (\ref{19})-(\ref{20}) we obtain the well known result:
\begin{equation} \label{50}
\Pi \left( {\bf x},t| \mathbf{0} \right) = W(t) \exp \left[ - \frac{ \left( \mathbf{x} - \mathbf{V} t  \right)^2 }{4 Dt} \right],
\end{equation}
where $ W(t) =  \left( 4 \pi D t \right)^{- 3/2} $ is a normalization factor, independent of the endpoints.

A more complex case consists of a Brownian particle immersed in a simple shear flow field, $ V_1 = \gamma x_2 $; $ V_2 = V_3 = 0 $. As shown in Mauri and Haber \cite{MH}, following the same procedure as before, we see that the minimum path equation (\ref{15}) reduces to:
\begin{equation} \label{52}
\ddot{y}_1 - \gamma \dot{y}_2 = 0; \hspace{2 em} \ddot{y}_2 + \gamma \dot{y}_1 - \gamma^2 y_2 = 0; \hspace{2 em} \ddot{y}_3 = 0,
\end{equation}
which, coupled to the conditions $ y_i(0) = 0 $; $ y_i (t) = x_i $, yields:
\begin{equation} \label{53}
y_1 (\tau) = C_1 \left( \widetilde{\tau}^3 - 6 \widetilde{\tau} \right) + C_2 \widetilde{\tau}^2; \hspace{2 em}  y_2 (\tau) = 3 C_1  \widetilde{\tau}^2 + 2 C_2 \widetilde{\tau}; \hspace{2 em}  y_3 (\tau) = Y_3 \tau / t,
\end{equation}
with $ \widetilde{\tau} = \gamma \tau $, where
\begin{equation} \label{54}
C_1 = \frac{\widetilde{t} x_2 - 2 x_1 }{\widetilde{t} \left( \widetilde{t}^2 + 12 \right) } ;  \hspace{2 em} C_2 = \frac{ \left( 6 - \widetilde{t}^2 \right) x_2 + 3 \widetilde{t} x_1}{\widetilde{t} \left(  \widetilde{t}^2 + 12 \right) },
\end{equation}
and $ \widetilde{t} = \gamma t $.  Substituting this result into Eqs. (\ref{19}) and (\ref{20}), we find the following Gaussian distribution:
\begin{equation} \label{55}
\Pi \left( {\bf x},t| \mathbf{0} \right) = W(t) \exp \left[ - \frac{3 \gamma \left( 2 x_1 - \widetilde{t} x_2 \right)^2 }{4 D \widetilde{t} \left( \widetilde{t}^2 + 12 \right)} - \frac{x_2^2 + x_3^2}{4 D t}  \right].
\end{equation}
Therefore, the variances of this distribution are:
\begin{equation} \label{PI.84}
\langle x_1^2 \rangle = 2 D t \left[ 1 + \frac13 \left( \gamma t \right)^2  \right];  \hspace{1 em} \langle x_1 x_2 \rangle = D \gamma t^2; \hspace{1 em} \langle x_2^2 \rangle = \langle x_3^2 \rangle = 2 D t.
\end{equation}

Here we see that, as expected, the mean freedisplace ment in the flow direction grows like $t^3$. Again, as expected, this result coincides with that by Foister and Van de Ven \cite{FV80} and by Katayama and Terauti \cite{KT96}, who obtained it by solving the Fokker-Planck and the Langevin equations, respectively.

%=================================================================
% References: Variant A
%=================================================================
% Back Matter (References and Notes)
%----------------------------------------------------------
% Style and layout of the references
\bibliographystyle{mdpi}

\begin{thebibliography}{99}
\bibitem{note1} Clearly, this is true only when the system is close to its equilibrium configuration, since, in general, phenomenological coefficients are far from being constant; for example, in the case of heat conduction, thermal conductivity, relating heat current and temperature gradient, should vary with temperature as $ T^{-2} $, which is very unlikely, and so the principle of minimum entropy production should be taken with great caution.
\bibitem{note2} This can also be proved explicitly, multiplying Eq. (\ref{15}) by $\dot{{\bf y}}$ and considering that $ \frac{d}{dt} = \frac{\partial}{\partial t} + \dot{{\bf y}} \cdot {\bf \nabla}$, obtaining $ d\mathcal{H}/dt = 0 $, i.e. $\mathcal{H}$ is constant, along the minimum path.
\end{thebibliography}
\makeatletter
\renewcommand\@biblabel[1]{#1. }
\makeatother

\end{document}